\documentclass[12pt]{iopart}
\usepackage{graphicx}
\begin{document}

\title{ 
Mixed Ising ferrimagnets with next-nearest neighbour
couplings on square lattices
}
\author{ W Selke$^1$ and C. Ekiz$^2$}
\address{$^1$ Institut f\"ur Theoretische Physik, RWTH Aachen, 52056
  Aachen, Germany}
\address{$^2$ Department of Physics, Adnan Menderes University,
  09010 Aydin, Turkey}
\date{\today}
\begin{abstract}
We study Ising ferrimagnets on square lattices with antiferromagnetic
exchange couplings between spins of values S=1/2 and S=1 on 
neighbouring sites, couplings between S=1 spins
at next--nearest neighbour sites of the lattice, and a 
single--site anisotropy term for the S=1 spins. Using 
mainly ground state considerations and extensive
Monte Carlo simulations, we investigate various aspects of
the phase diagram, including compensation points, critical
properties, and temperature dependent anomalies. In contrast to previous
belief, the next--nearest neighbour couplings, when being of
antiferromagnetic type, may lead to compensation points.

\end{abstract}
\submitted{\JPCM}

\section{Introduction}
\label{sec1}

Mixed spin ferrimagnetic Ising models have been studied
for some time, with a renewed interest, especially, in 
connection with 'compensation
points', see, for instance, [1-11]. These 
points occur at temperatures, below the critical temperature, at which the
sublattice magnetizations cancel
exactly, giving zero total moment. As the temperature is tuned through
such a point the total magnetization changes sign, which may be
used in technological applications, most notably in
magnetic recording.

One of the simplest such models consists of classical Ising spins
S=1/2 and S=1 on a square or simple cubic lattice with the spins
of different type being located on neighbouring
sites of the lattice. Its Hamiltonian
may be written in the form

\begin{equation}
{\cal H} = J_1 \sum_{\langle i,j\rangle} {\sigma_i S_j} + D \sum_{j\in B} {S_j}^2
\end{equation}

\noindent
where $J_1 > 0$ denotes the antiferromagnetic coupling between 
spins $\sigma_i = \pm 1/2$ on the sites of sublattice 'A', and neighbouring 
spins $S_j = 1,0,-1$ on sites forming the sublattice 'B'. Then, spins
on each sublattice tend to order ferromagnetically, with opposite
sign for the two types of spins. $D$ is the
strength of a single--site anisotropy (or crystal--field) term acting
only on the S=1 spins of sublattice B.

One may also choose $\sigma_i = \pm 1$ rather than $\pm 1/2$. The
change has to be taken into account when calculating sublattice
magnetizations and defining the compensation
point. Otherwise, the modified convention simply amounts to a rescaling of
the exchange coupling \cite{bue,seloit}.

The mixed spin Ising model, eq. (1), is known, see,
e.g., \cite{zha,bue,seloit}, to
lead to compensation points for
simple cubic lattices, in accordance
with mean--field theory \cite{kan}, but not for square
lattices, in contrast to mean--field theory. As has been
observed already some years ago by Buendia and
Novotny \cite{bue,bue2}, compensation
points may occur on square lattices, when adding a (ferromagnetic)
coupling between next--nearest neighbouring (nnn) spins on
the A sublattice. On the other hand, the authors
did not find any evidence for compensation points, when considering
nnn couplings for B spins, instead of the ones for A spins.

In the following article we shall challenge the latter
suggestion, which seems to have been taken
for granted by others, see, e.g., \cite{god}. Indeed, based
on extensive Monte Carlo (MC) simulations, we shall present
clear evidence for compensation points due to nnn antiferromagnetic
interactions between B, or S=1, spins on the square lattice. In fact, the
effect is not easy to identify. 

The outline of the article is as follows. In section 2, to set
the scene for the following parts, we define the
model with nnn couplings between S=1 spins, discuss its ground
states, and describe details of the simulations. The
following section deals with the compensation points. In
section 4, MC results on critical properties and
temperature dependent anomalies of the model will be presented. A
brief summary is given in the final section.

\section{Model, ground states, and simulations}
\label{sec2}
We shall study the following mixed spin Ising model on
a square lattice

\begin{equation}
{\cal H} = J_1 \sum_{\langle i,j\rangle} {\sigma_i S_j} + D \sum_{j\in B} {S_j}^2
- J_2 \sum_{\langle i,j\rangle\in B} {S_i S_j}
\end{equation}

\noindent
with antiferromagnetic nearest neighbour couplings, $J_1>0$ between
A spins, $\sigma_i =\pm 1$, and B spins, $S_j= 0, \pm 1$, a
crystal--field term of strength $D$ acting on the B spins, and
nnn couplings, $J_2$, between B spins.  Note that we set 
A spins equal to $\pm 1$, in accordance with previous
work \cite{bue,seloit}. To identify possible
compensation points, care is then needed in defining the
magnetization of the sublattice A, see above and below.
 
To determine the ground states of the model in the
($D/J_1$,$J_2/J_1$) plane, we first select, like
before \cite{bue,bue2,bue3}, the structures which are
stable among the ones
described by 2$\times$2 cells of spins on the square lattice. One 
readily observes the possibility of degenerate ground states,
eventually with indefinitely large unit cells. Finally, we check
the tentative ground state structures by
monitoring spin configurations at very low temperatures obtained from
careful cooling MC runs for lattices of various sizes. The 
resulting ground state phase diagram is depicted in figure 1.    

The phase diagram at vanishing temperature, $T=0$, comprises
four structures. The antiferromagnetic (AF) structure is stable for
ferromagnetic or weakly antiferromagnetic nnn couplings, $J_2$, and
negative or sufficiently small positive values of the crystal
field, $D$. The spins on each of the
two sublattices order ferromagnetically, $\pm 1$, with opposite sign   
on the sublattices A and B.

In two of the remaining other three
ground states, S=1 spins may be in the state 0,
due to the single--site anisotropy term, $D$. Actually, for 
sufficiently large values of 
$D$, all spins of the sublattice B are in state 0. We
abbreviate the resulting structure as '0II'. The structure is highly
degenerate, with each A spin being either $-1$ or 1, leading
to a $2^{N_A}$--fold degeneracy for $N_A$ sites on the
sublattice A. In contrast, in the '0I' structure, see figure 1, only
half of the S=1 spins are in the state 0. They form, on the
sublattice B, in the standard Wood's notation for
overlayers \cite{oura}, a periodic c($2\times 2$) superlattice. The other half
of the spins on the B sites are ferromagnetically ordered, $\pm 1$. The  
spins on the sublattice A are ferromagnetically ordered as well, with
opposite sign, compared to that of the B spins.

\begin{figure} [h]
\vspace{1cm}
\begin{center}
 \includegraphics[width=0.55\linewidth]{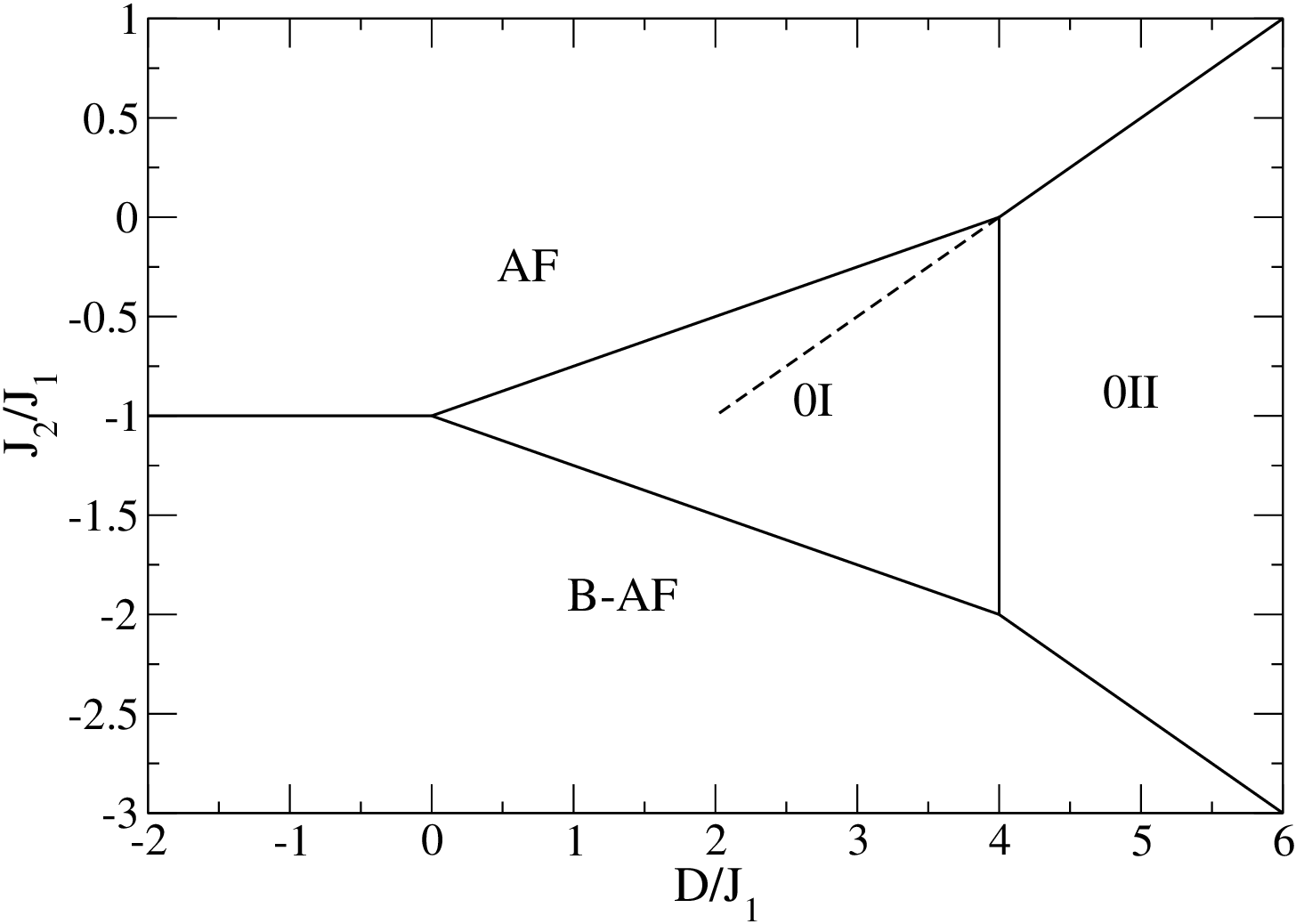}
\end{center}
\label{fig1}
\vspace{-0.5cm}
\caption{Ground state phase diagram of the mixed spin model, eq. (2),
  on a square lattice, with solid lines separating the
  different phases. The dashed line in the 0I phase refers to
  equation (7), see text.} 
\end{figure}

In the fourth ground state structure, all spins on the sublattice B are
antiferromagnetically aligned, due to the dominant antiferromagnetic nnn
coupling, J$_2$. The structure may be abbreviated as B--AF. Each spin on
the sublattice A may be either $-1$ or 1, leading, as for 0II, to
a large degeneracy. 

Note that there may be additional degeneracies at
borderlines between different ground state structures \cite{seloit}.

To determine thermal properties of the model, (2), we mainly perform
MC simulations, using the Metropolis algorithm with
single--spin flips, providing, indeed, the required
accuracy,  so that there is no need to apply other techniques
like cluster--updates or the Wang--Landau approach \cite{bil}. We consider
lattices with $L$x$L$ sites, employing full
periodic boundary conditions. $L$ ranges 
from 4 to 120, to study finite--size effects. Typically, we do runs
of $5\times 10^6$ to $10^7$ Monte Carlo steps per spin, where
averages and error bars may be obtained from evaluating a few of such
runs, using different random numbers. These quite long runs lead to
reliable data, as before \cite{seloit}. The estimated errors are
usually smaller than the sizes of the symbols in the
figures, and they are shown there only in a few cases.

We record the energy per site, $E$, the
specific heat, $C$, both from the energy fluctuations 
and from differentiating $E$ with respect to the temperature, and
the absolute values of the sublattice magnetizations
of the two sublattices

\begin{equation}
|m_A| = <|\sum_{A} {\sigma_i}|>/(2 (L^2/2))
\end{equation}

\noindent
and 

\begin{equation}
|m_B| = <|\sum_{B} {S_j}|>/(L^2/2)
\end{equation}

\noindent
as well as the absolute value of the staggered magnetization of
sublattice B, describing the ordering for the B--AF structure,

\begin{equation}
|m^{st}_B| = <|\sum_{B^+}{S_i}- \sum_{B^-}{S_j}|>/(L^2/2)
\end{equation}

\noindent
with B$^{+,-}$ denoting an obvious bipartition of
sublattice B. The brackets $< >$ denote the thermal average. Note the factor
of 1/2 in the definition of $|m_A|$, taking into account the
correct length of the S=1/2 spins, so that $|m_A(T=0)|= 1/2$ for 
the ferromagnetic ground state of sublattice A. In
addition, the corresponding sublattice (staggered) 
susceptibilities, $\chi_A$, $\chi_B$, and $\chi_B^{st}$, have been computed
from the fluctuations of the (staggered) sublattice magnetizations. We
also analyse the fourth--order cumulant of various (sublattice) order
parameters, the Binder cumulant \cite{bin}, defined by
 
 \begin{equation}
  U = 1- <m^4>/(3 <m^2>^2)
\end{equation}

\noindent
with $<m^2>$ and $<m^4>$ being the second and fourth moment of 
(staggered) magnetizations of sublattice A or B. Finally, we monitor
typical equilibrium Monte Carlo configurations, illustrating the microscopic
behaviour of the system and providing, e.g., evidence for the ground states
structures, as mentioned above.

To test the accuracy of the simulations, we compared our MC data to
those of previous accurate numerical work and to
exact results by enumerating all possible configurations for small
lattices with $L= 4$ \cite{bue,seloit}.

\section{Compensation points}
\label{sec3}

Compensation points occur at temperatures $T_{comp}$, where
both sublattice magnetizations cancel each other, with
a vanishing total magnetization. In finite systems, as
one is studying in MC simulations, a convenient
and efficient way \cite{bue} to locate such points is to use the
crossing condition $|m_A|(T_{comp})= |m_B|(T_{comp})$. At
low temperatures sufficiently far below the phase transition, finite--size 
effects are expected to be very weak, because there the compensation 
effect is not related to critical phenomena. 

\begin{figure} [h]
\vspace{1cm}
\begin{center}
\includegraphics[width=0.55\linewidth] {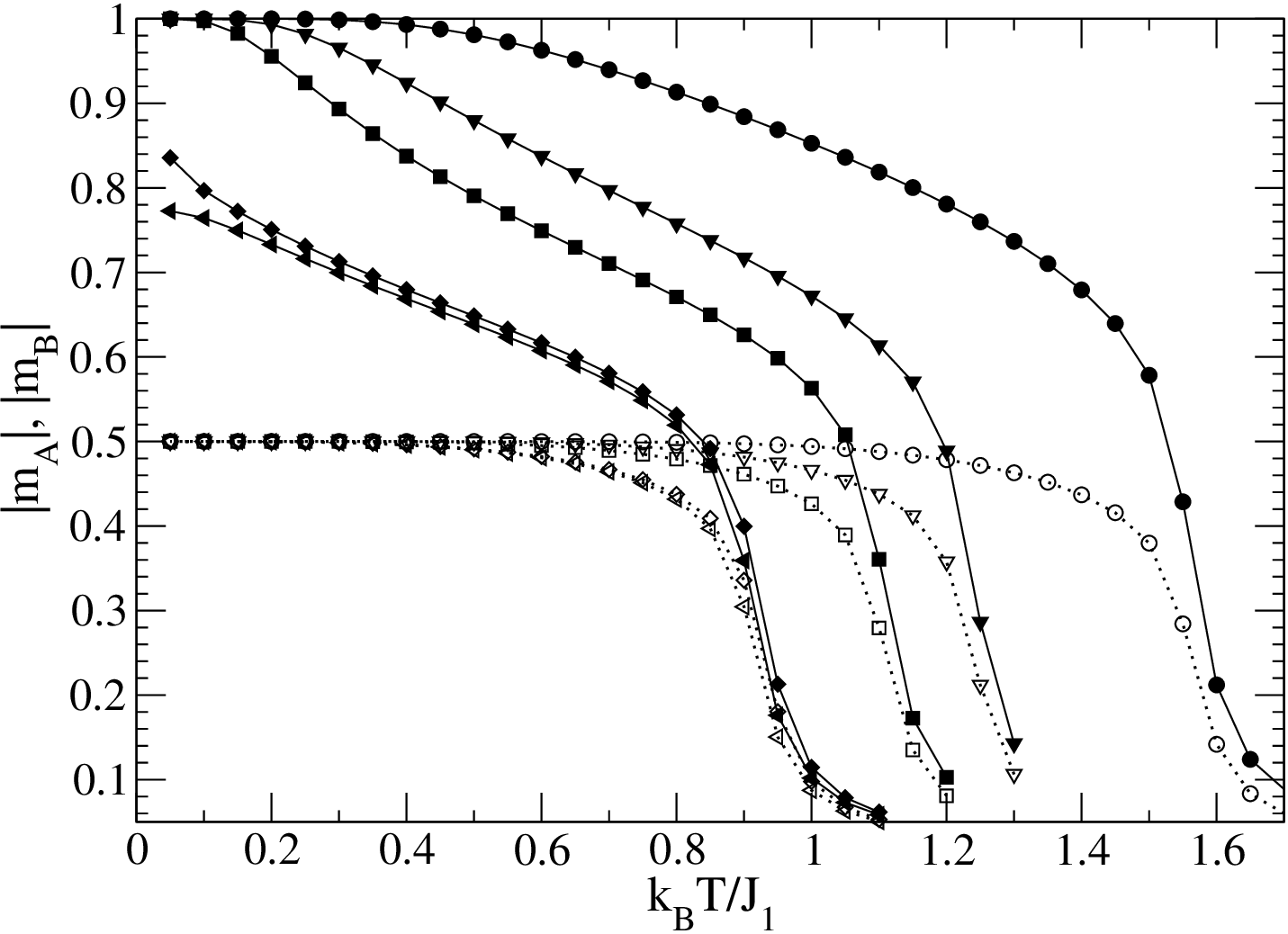}
\end{center}
\label{fig2}
\caption{Sublattice magnetizations $|m_A|$ (open symbols, dotted
    lines) and $|m_B|$ 
    (full symbols, solid lines) at fixed
    $D/J= 3.0$ and varying $J_2/J_1$= 0.25 (circles), 0.0 (triangles
    down), $-0.1$ (squares), $-0.24$ (diamonds), and $-0.26$ (triangles
    left). Lattices of size $60^2$ are simulated.} 
\end{figure}

In the AF ground state, one has
$|m_B|(T=0)$=1, while $|m_A|(T=0)$= 1/2. Now, at non--zero
temperatures, $|m_B|(T)$ may fall off quite rapidly due to 
antiferromagnetic couplings $J_2$ and due to a relatively large
single--site anisotropy term, $D$, which favours flips of
S=1 spins to state 0. Furthermore, at zero temperature, $|m_B|(T=0)$ drops
from 1 to 1/2, when passing through the borderline between the AF and
0I ground states. Accordingly, one might speculate that compensation
points, at low temperatures in the AF phase, may show up close
to the AF--0I borderline at $T$=0. In
fact, our preliminary mean--field calculations seem to suggest
that lines of compensation points may spring from this AF--0I borderline. 

However, in our MC simulations, investigating carefully
several cases, we find no evidence for such
compensation points in the AF phase. An
example is depicted in figure 2. There, $D/J_1$  is fixed at
3.0, and $J_2/J_1$ is varied, from +0.25 to $-0.26$, so that
the AF--0I border at $J_2/J_1$ = $-0.25$ is approached and
crossed. One observes, that $|m_B|$ is always larger than $|m_A|$, albeit 
the difference between the two sublattice magnetizations may get quite small
when decreasing $J_2$. MC data for lattices of fixed size, $L= 60$, 
are displayed. The observation on the absence of
compensation points in the AF phase also holds for other lattice
sizes as well as for other values of $J_2/J_1$ and $D/J_1$.

\begin{figure} [h]
\vspace{1cm}
\begin{center}
\includegraphics[width=0.55\linewidth] {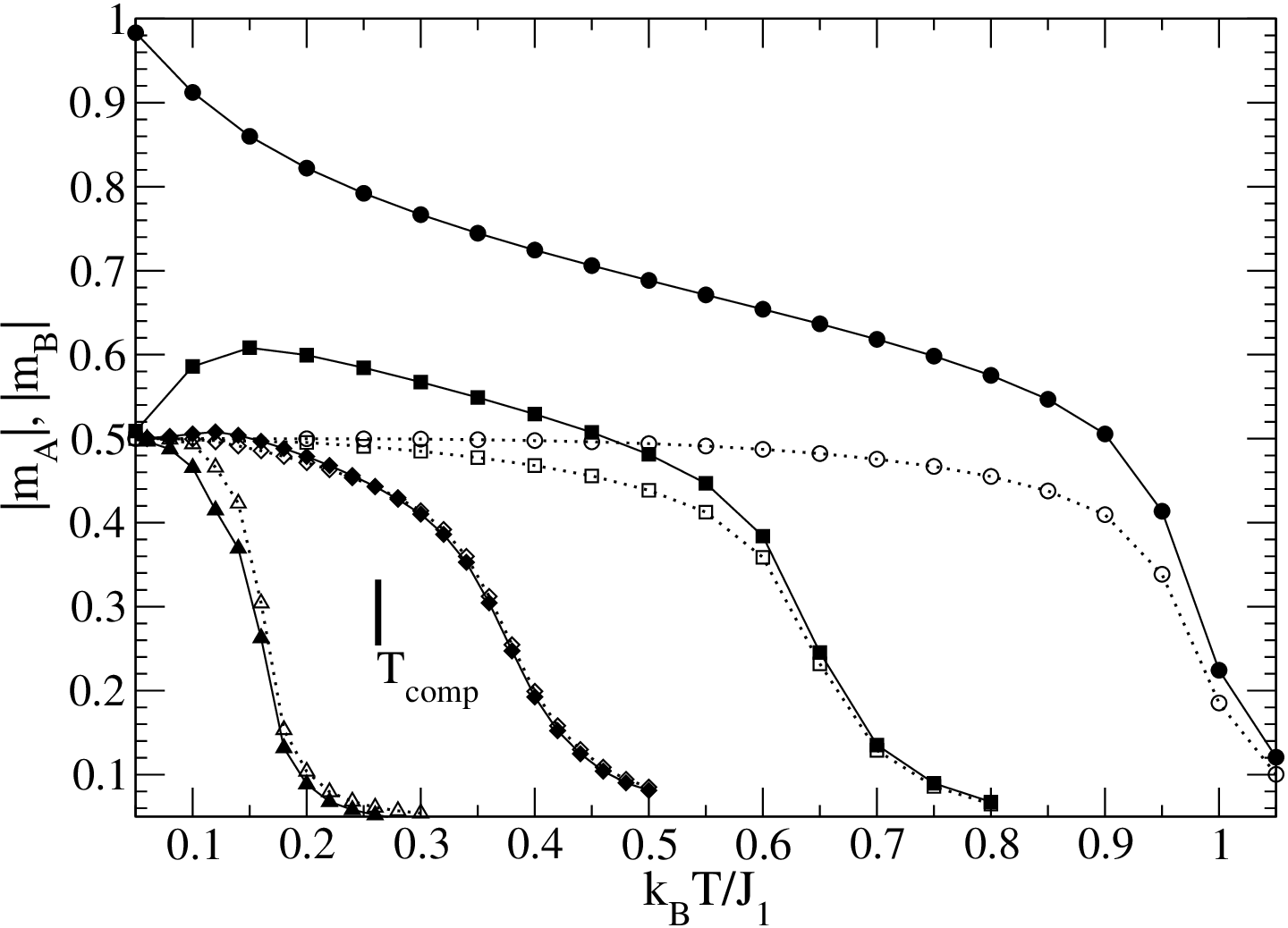}
\end{center}
\label{fig3}
\caption{Sublattice magnetizations $|m_A|$ (open symbols, dotted
    lines) and $|m_B|$
    (full symbols, solid lines) at fixed $J_2/J_1 = -0.2$, and varying
    $D/J_1$= 3.0 (circles), 3.4 (squares), 3.57 (diamonds), and
    3.7 (triangles up), simulating lattices of size $L$= 60. At
    $D/J_1$ =3.57, a compensation point shows up, $T_{comp}$.} 
\end{figure}

On the other hand, compensation points will be argued below
to occur in the 
0I phase, for $J_2/J_1 > -1.0$, arising at vanishing temperature 
from the line

 \begin{equation}
  J_2/J_1= -2 + D/(2J_1)
\end{equation}

\noindent
which is depicted as the dashed line in the ground state phase
diagram, figure 1.

Let us first present numerical support for this claim, followed 
then by low temperature energy considerations backing it up. A
numerical example is shown in figure 3, where, for lattices with
$60^2$ sites, the temperature dependences of the sublattice
magnetizations, $|m_A|$ and $|m_B|$, are shown at fixed
nnn coupling, $J_2/J_1 =-0.2$, and varying the strength
of the single-site anisotropy term, $D/J_1$, from 3.0 to 3.7. Note
that for this ratio of couplings, at $T= 0$, the AF--0I border is at 3.2,
and the origin, at vanishing temperature, of the line of compensation
points would be, according to equation (7), at $D/J$= 3.6. As discussed
in the context of figure 2, one observes that $|m_B|(T)$ is 
always larger than $|m_A|(T)$ when being in the AF phase. However, in
the 0I phase, here, at $D/J_1$= 3.57, there seems to be 
a compensation point, with $k_BT_{comp}/J_1 \approx 0.26$, clearly
below the phase transition: At
lower temperatures, $|m_B|$ is still larger than $|m_A|$, with
reverse ordering at $T > T_{comp}$. By further increase of
$D$, $D/J_1$= 3.7, $|m_A|(T)$ is larger than $|m_B|(T)$, at all temperatures
up to $T_c$.

 \begin{figure} [h]
\vspace{1cm}
\begin{center}
\includegraphics[width=0.55\linewidth]{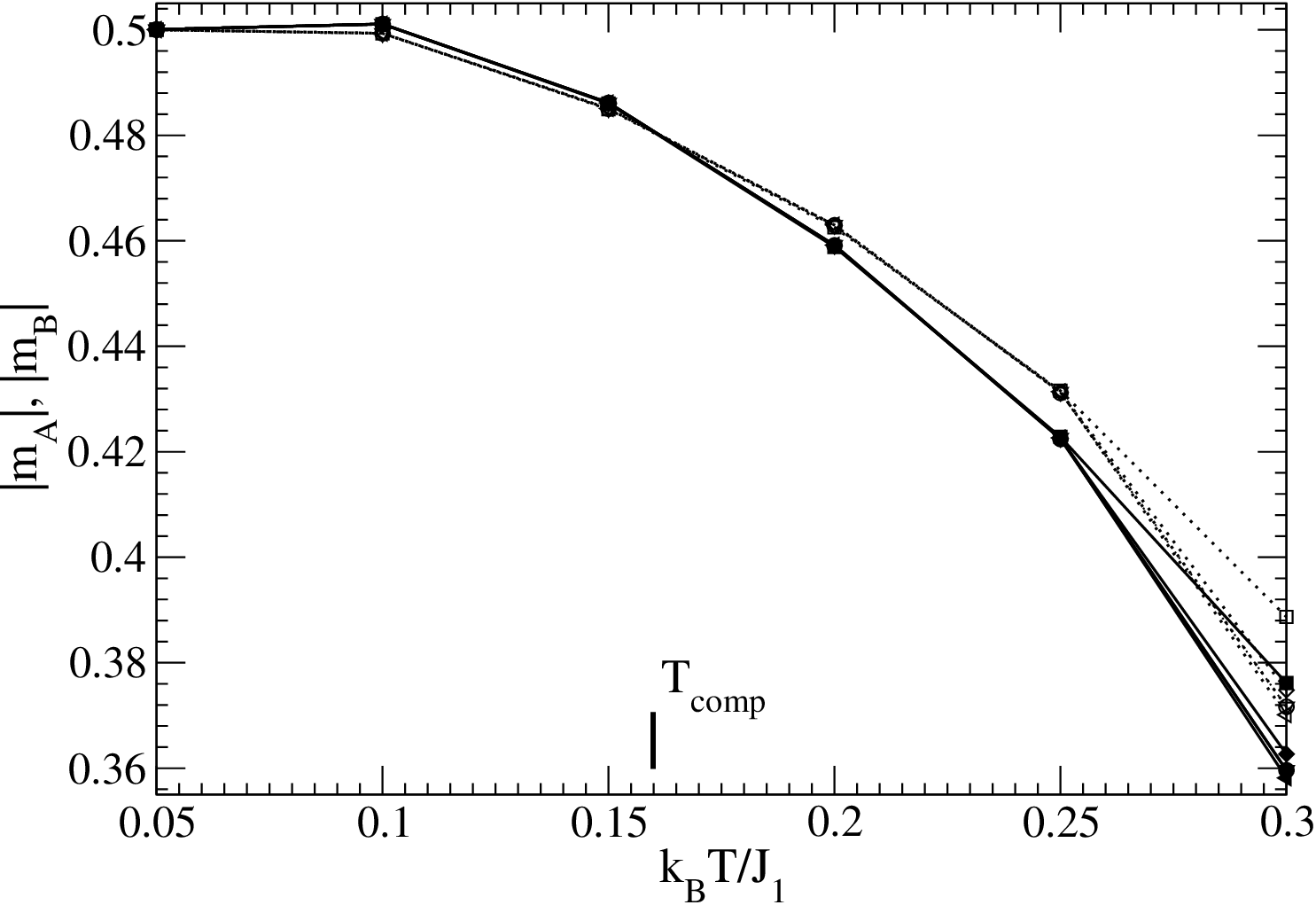}
\end{center}
\label{fig4}
\caption{Sublattice magnetizations $|m_A|$ (open symbols, dotted
    lines) and $|m_B|$
    (full symbols, solid lines) at $J_2/J_1 = -0.2$ and $D/J_1$= 3.59, with
    lattice sizes $L$= 20 (squares), 40 (diamonds), 60 (triangles left),
    80 (triangles down), and 120 (circles). The compensation point
    is located at $k_BT/J_1 \approx 0.16$.}  
\end{figure}
 
To establish numerically the presence of compensation points, careful
finite--size analyses are required, as illustrated
in figure 4. Here, MC data for various lattices sizes, with $L$ 
ranging from 20 to 120, are displayed, at $J_2/J_1= -0.2$ and 
$D/J_1$= 3.59, i.e. very close to the point, from which, according to 
(7), the line of compensation points may arise. In fact, a compensation
point may be located at 
$k_BT_{comp}/J_1 \approx 0.16$. Below that temperature, $|m_B|(T)$
supercedes $|m_A|(T)$, with finite--size dependences being
extremely small. Then, at $T_{comp}$, a crossing of the
two sublattice magnetizations occurs. The finite size effects are
still very weak up to $k_BT/J_1 \approx 0.25$, providing
clear evidence on the existence of the compensation point
in the thermodynamic limit. 

Indeed, similar finite-size analyses
allow one to locate the rather steeply rising line of compensation points, at
fixed nnn couplings $J_2/J_1= -0.2$. Results on the compensation
points, $T_{comp}$, are summarized in figure 5, together with
estimates for the transition line, $T_c$, to the paramagnetic
phase.

\begin{figure} [h]
\vspace{1cm}
\begin{center}
\includegraphics[width=0.55\linewidth] {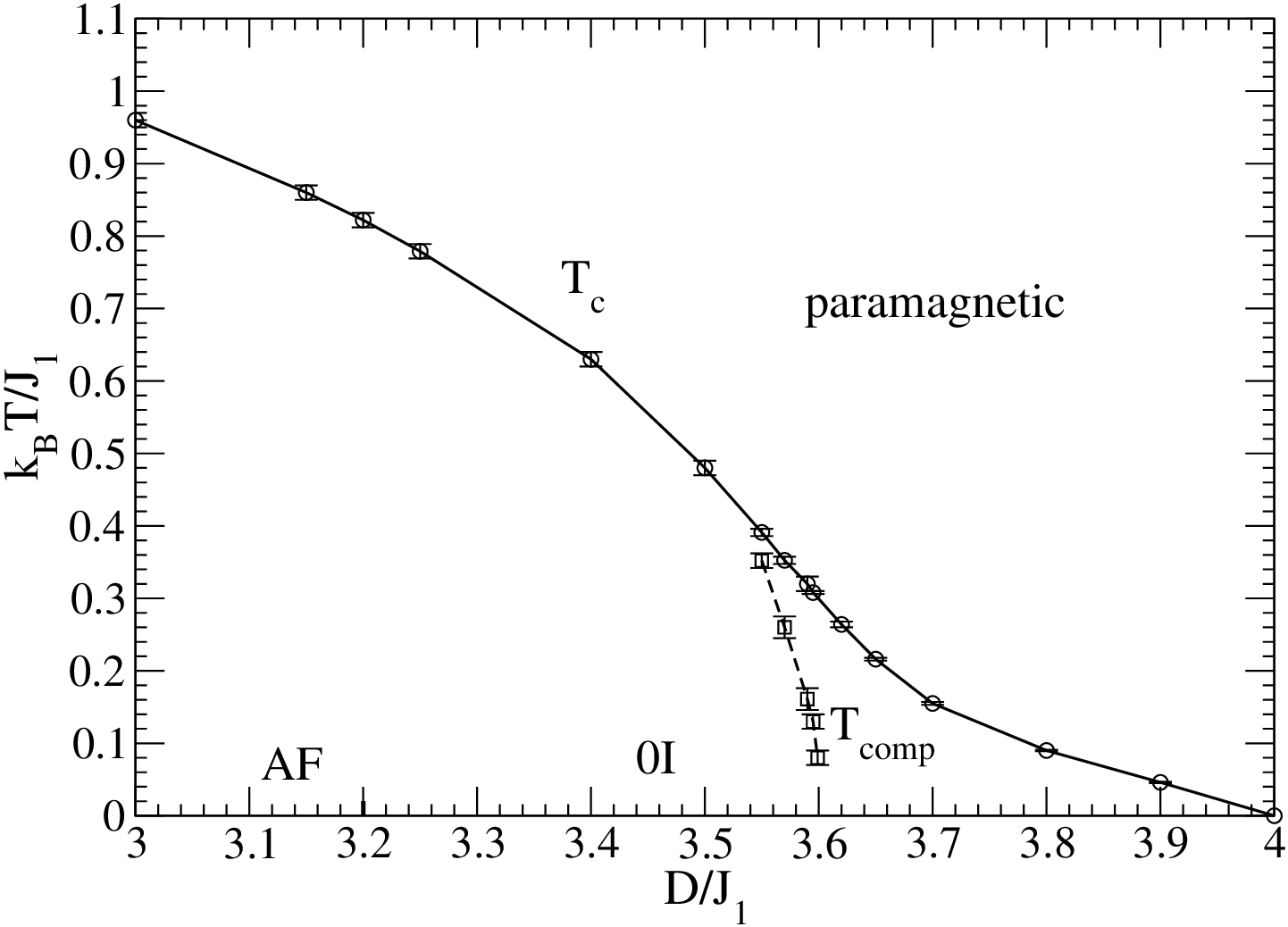}
\end{center}
\label{fig5}
\caption{Phase boundary to the paramagnetic phase, $T_c$ (solid line), and
    compensation points, $T_{comp}$ (dashed line), at $J_2/J_1 =
    -0.2$, varying $D/J_1$.}
\end{figure}

The critical line, $T_c$, has been determined by monitoring
the size dependent positions of maxima in the
specific heat $C$, in the susceptibility of the sublattice A, and 
in the intersection points of the Binder cumulant, especially, of
sublattice A. The
intersections of the Binder cumulant $U(L,T)$ for
successive lattice sizes \cite{bin}
turn out to have the smallest finite--size effects, thereby being
most efficient in estimating the critical temperature.

The line of compensation points seems to
start at zero temperature at $D/J_1$= 3.6 for
$J_2/J_1= -0.2$, in agreement with equation (7). Furthermore, it
extends only over quite a small
region, with the compensation point coinciding with the
critical point at about $D/J_1= 3.52 \pm 0.02$. 

\begin{figure} [h]
\vspace{1cm}
\begin{center}
\includegraphics[width=0.55\linewidth] {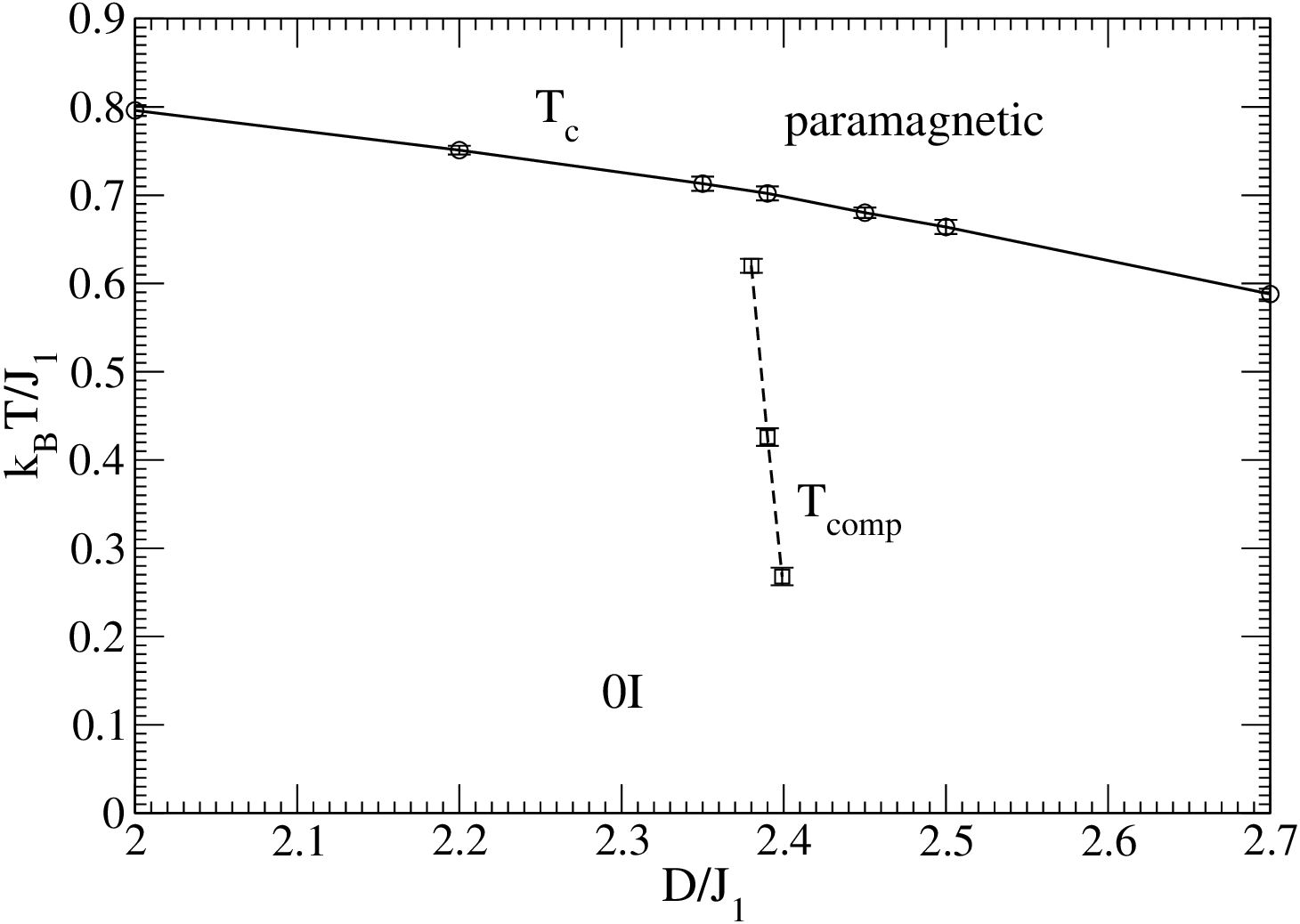}
\end{center}
\label{fig6}
\caption{Phase boundary to the paramagnetic phase, $T_c$ (solid line), and
    compensation points, $T_{comp}$ (dashed line), at $J_2/J_1 =
    -0.8$, varying $D/J_1$.}  
\end{figure}

Let us now turn to the low temperature considerations leading to
(7). Specifically, we consider, for the 0I structure, the
one--spin flip excitations on sublattice B. For
concreteness, we analyse ground states with spins being in state $-1$
or 0 on sublattice B, and in state 1 on sublattice A. Then there are
four possible flips of S=1 spins, namely flipping a spin from its
state 0 to the state (i) $-1$ or (ii) 1, or flipping a spin from its 
state $-1$ to state (iii) 0 or (iv) $+1$. Obviously, only flip (i)
will increase the sublattice magnetization $|m_B|$ above its value at 
zero temperature, $|m_B|(T=0) =1/2$, while otherwise $|m_B|(T)$
will have a  negative slope at low temperatures. Now, one
may readily calculate the energies
of all four elementary flips. One obtains that
flip (i) costs the lowest
energy either followed by flip (iii), when $D/J_1 < 2J_2/J_1 +4$,  or
followed by flip (ii), when  $J_2/J_1 > -1.0$. Equation (7) is then
obtained under the assumption, that an initial
increase of $|m_B(T)|$ is needed to have a compensation
point, with $|m_B| > |m_A|$ at sufficiently low
temperatures.  Otherwise, in the 0I phase, $|m_A| > |m_B|$ at all temperatures
$0 < T < T_c$, excluding a compensation point.

We further checked numerically the hypothesis, (7), in determining
lines of compensation points by fixing the nnn coupling $J_2/J_1$ not
only at $-0.2$, but also at $-0.4$ and $-0.8$, varying the strength
of the single--site anisotropy term, $D/J_1$. Indeed, we observe compensation
points in the 0I structure, which seem to arise, at zero
temperature, from $D/J_1$= 3.2 at $J_2/J_1= -0.4$ and
from $D/J_1$= 2.4 at $J_2/J_1= -0.8$, in agreement with
(7). The latter case is displayed in figure 6.

Note that the range of values of $D/J_1$, in which there
is a line of compensation points, shrinks appreciably as one decreases
$J_2/J_1$. This may be seen by comparing figures 5 and 6. Thence, it
becomes increasingly difficult to locate compensation points for
lower values of $J_2/J_1$.

\section{Phase transitions and anomalies}
\label{sec4}

While mixed spin Ising models are of much interest because
of the potential presence of compensations points, they may exhibit
other intriguing thermal properties as well.

We shall focus on two aspects, the characterization of the phase
transitions associated with the various ground states, and
anomalies in the temperature dependence, especially, of 
the sublattice magnetization and the specifc heat $C$.

One expects that there is no phase transition associated with
the highly degenerate 0II ground state for the following reasons: At
zero temperature, all spins
of sublattice B are in state 0, while each spin of
sublattice A may be either $-1$ or 1. The single--site anisotropy
term, favouring the spin state 0 on sublattice B, does not
support long--range order, in close analogy to the situation
in the Blume--Capel model \cite{blucap}. The spins on
sublattice A act merely like a fluctuating random field, and
may not lead to a phase transition neither. In fact, our MC
data, for various quantities, like the specific heat $C$ or
the probability to encounter a spin in state 0, give no indication
for singular thermal behaviour.

\begin{figure} [h]
\vspace{1cm}
\begin{center}
\includegraphics[width=0.55\linewidth] {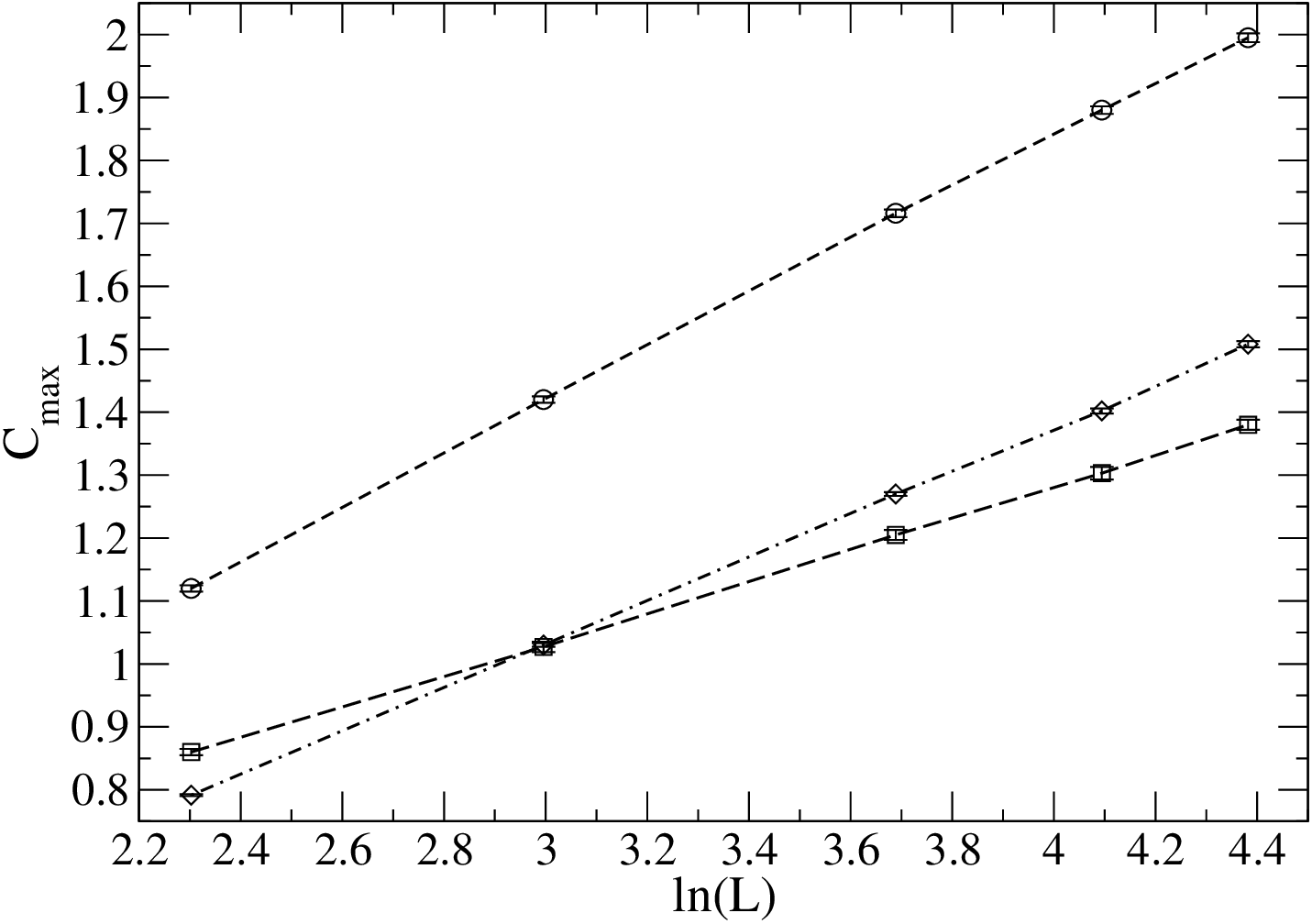}
\end{center}
\label{fig7}
\caption {Height of the critical maximum of the
     specific heat, $C_{max}$ versus logarithm of the 
     lattice size, $\ln L$, for $L$= 10, 20, 40, 60, and
     80 at (a) (circles) $J_2/J_1 =-0.3$,
     $D/J_1$ = 1.5, (b) (squares) $J_2/J_1 = -1.0$, $D/J_1$= 1.5, and
     (c) (diamonds) $J_2/J_1 = -1.5$, $D/J_1$= 2.5. The three cases
     refer to the (a) AF, (b) 0I, and (c) B--AF structures.} 
\end{figure}

In case of the other three ground states, AF, 0I, and B--AF, we
determine the universality
class \cite{fish,pel} by analysing specific heat and susceptibilities. 
In particular, we monitor the size, $L$, dependence of that maximum 
of the specific heat, $C_{max}(L)$, which goes over into a singularity
in the thermodynamic limit, $L \rightarrow \infty$ (note that there
may be other maxima as will be discussed below in the context of
anomalies). Furthermore, we study the size dependence of
the maxima in (sublattice) susceptibilities. For
the AF and 0I structures, we focus on $\chi_{A,max}(L)$. For
the B--AF structure, we record the staggered susceptibility of
the antiferromagnetically ordered sublattice B, $\chi^{st}_{B,max}(L)$. 
In all three cases, the critical behaviour is found to be
consistent with having
phase transitions in the universality class of the standard
two--dimensional Ising model.

Typical MC results on the size, $L$, dependence of maxima in the
specific heat, $C_{max}$, are depicted in
figure 7. For all three types of ground states, AF, 0I
as well as B--AF, one approaches to a good degree, already for 
lattices of moderate sizes, the
form $C_{max} \propto \ln L$, being characteristic for the
two--dimensional Ising universality class. Note that the prefactor in
front of the logarithmic term depends strongly on $J_2/J_1$ and
$D/J_1$. Certainly, such Ising--like
critical behaviour may be expected in the AF case. In case of the B-AF
phase, the spins on sublattice B order antiferromagnetically, with the
A spins providing effectively a randomly fluctuating field, which may
be argued to be irrelevant for the universality class. Finally, in the
0I case, the Ising--like criticality may be understood
by the ferromagnetic order of the spins on sublattice A.

\begin{figure} [h]
\vspace{1cm}
\begin{center}
\includegraphics[width=0.55\linewidth] {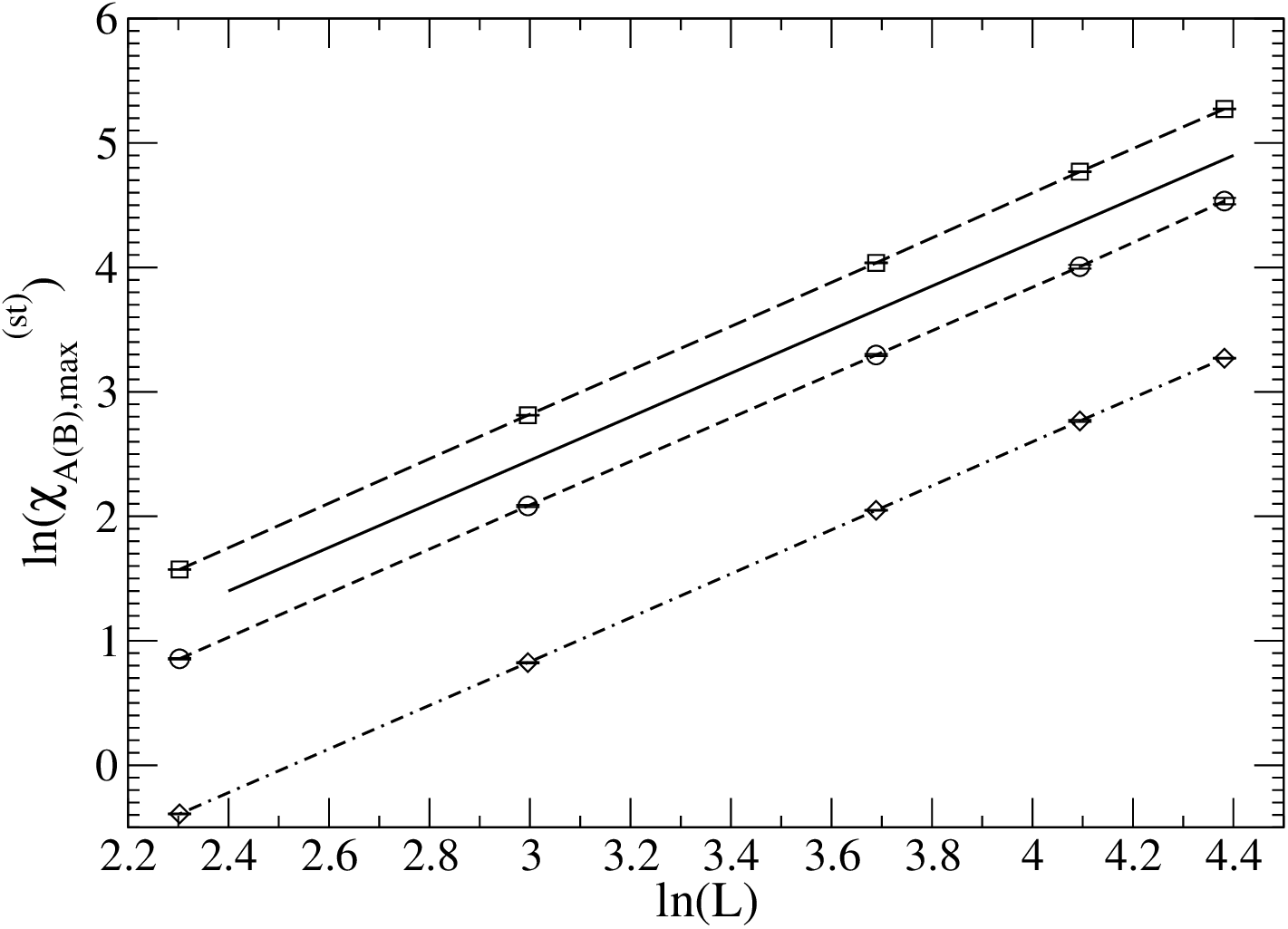}
\end{center}
\label{fig8}
\caption{Log--log plot for maxima of (staggered) sublattice susceptibilities
 $\chi_{max}$ versus lattice size $L$ for the AF, 0I, and B--AF
 cases, with the same parameters and notation as in figure
 7. In case of AF and 0I structures, $\chi_{A,max}$, in
 case of the B--AF structure, $\chi_{B,max}^{st}$ is recorded. For
 comparison, the solid line shows $\chi_{max} \propto L^{7/4}$.} 
\end{figure}

The analysis of the susceptibilies confirms the findings on
the specific heat. Typical examples, for the same model
parameters as in figure 7, are shown in figure 8. The size 
dependent maxima in the susceptibilities are observed to follow the
form $\chi_{max} \propto L^{7/4}$, with rather small
corrections, for all three cases, AF, 0I, and B--AF. This
behaviour may be quantified by calculating the slope between
successive points in the doubly logarithmic plot of the MC data. The
resulting effective local critical exponent
of the susceptibility is near 7/4, even for moderate 
lattice sizes. Thence, in all
three cases, criticality seems to belong to the two--dimensional
Ising universality class.
  
Care is needed when attempting to determine the universality class 
from the value of the critical Binder cumulant
$U^*$= $U(T_c,L \longrightarrow \infty)$, because that
value is known to depend on boundary conditions, shape, and
anisotropy of the correlations \cite{bin,doh,sel}. Note that in
the cases we studied, its value seems to be close to that of the
standard two--dimensional Ising model with periodic boundary
conditions for lattices of
square shape, $U^* \approx 0.6107$ \cite{kamin}. However, the
dependence, especially on anisotropy, may be
very weak \cite{doh,sel}, and a reliable analysis may require
an exact or, at least, extremely
accurate determination of the critical point. We therefore
refrained from such an analysis here. 

\begin{figure} [h]
\vspace{1cm}
\begin{center}
\includegraphics[width=0.55\linewidth] {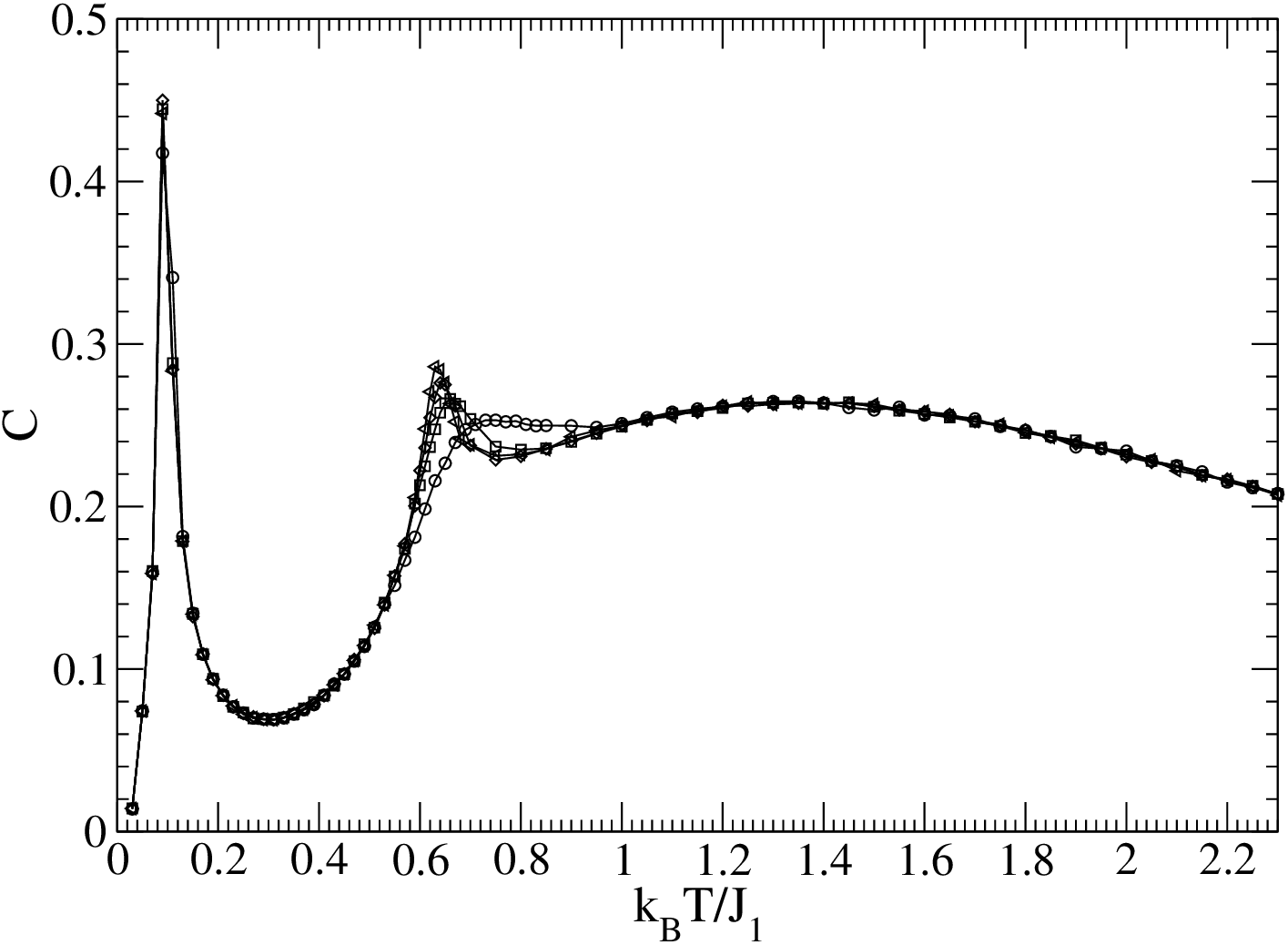}
\end{center}
\label{fig9}
\caption{Specific heat $C$ as a function of temperature, $k_BT/J_1$, at
    $J_2/J_1 =-0.2$ and $D/J= 3.4$, simulating lattices of sizes
    $L$= 20 (circles), 40 (squares), 60 (diamonds), and 80 (triangles
    left).} 
\end{figure}

Let us now turn to discussing anomalies, exhibiting intriguing
non--monotonic temperature dependences, which
are not related to phase transitions. Especially, we
monitor the magnetizations of sublattice B, $|m_B|$, the specific
heat $C$, and the thermally averaged occupancy of B sites with spins
in the state 0, $p_B(0)$.
 
An example for such anomalies is the overshooting of $|m_B|(T)$ at
low temperatures in the vicinity of the 
0I ground state, illustrated in figures 2 and 3. As discussed
above, the overshooting is related to decreasing the number of spins in
state 0, by flipping such spins
to, say, $-1$, costing minimal energy. Indeed, the anomaly
in $|m_B|$ is monitored to be accompanied by a pronounced
decrease in $p_B(0)$. 

Anomalies may also occur in the specific heat, being
caused either by a rapid increase or decrease in the number
of S=1 spins in state 0. For instance, such an anomaly in $C$ has
been observed before \cite{seloit} in the AF phase 
at $J_2/J_1= 0$, close to the 0II structure. There, monitoring
$C(T)$, a
three--peak--structure of the specific heat has been found, with a
non-critical maximum at low temperatures, due to easy flips of spins
in the state 0, followed, at higher temperature, by a strongly
size dependent critical peak, and, at even larger
temperature, another non--critical maximum, due to flipping
of single spins for sublattices with rather large clusters
of '+' or '$-$' spins.

We observe a similar thermal behaviour of the specific heat in
the 0I phase, when turning on the nnn coupling, $J_2 <0$. An
example is depicted in figure 9, at $J_2/J_1= -0.2$ and
$D/J_1$= 3.4. However, here the peak at lowest
temperature, $T_l$, is due to a rather drastic decrease
of B spins in state 0, which has been discussed above. Actually, by 
enhancing the strength of the single--site anisotropy
term, $D/J_1= 3.6$, the 
position of that peak, at $T_l$, shifts first to somewhat higher
temperatures. It still corresponds to a
lowering, with temperature, of the average number
of 0's for the sublattice B, $p_B(0)(T)$. When moving
even closer to the 0II structure, $D/J_1$ = 3.8 and 3.9, $T_l$ tends to be 
shifted towards lower temperatures with increasing $D$. The peak
position seems to follow the
same dependence as for vanishing $J_2$ \cite{seloit}, namely
$k_BT_l/J_1 \propto 4-D/J_1$. As may be seen from monitoring $p_B(0)$, the
peak is then, as in the limit $J_2= 0$, due to an increase in the number
of B spins in state 0 at low temperatures.

\section{Summary}
\label{sec5}

We have studied a mixed spin Ising model with
antiferromagnetic couplings, $J_1$, between spins S= 1/2 and S= 1
on neighbouring sites of a square lattice, augmented
by couplings, $J_2$, between spins S= 1 on next--nearest neighbouring sites
of the lattice. An additional quadratic single--site
anisotropy term, $D$, acts upon the S=1 spins. Based mainly on
ground state considerations and on extensive
standard Monte Carlo simulations, we have determined the
ground state phase diagram in the ($D/J_1,J_2/J_1$)
plane and identified compensation points, types of 
phase transitions corresponding to different ground state
structures as well as anomalies for various physical quantities. 

In particular, compensation points are found to exist, in
contrast to previous belief. They exist for antiferromagnetic nnn
couplings, $J_2>-1.0$, in the 0I phase, springing, at
zero temperature, from the line $J_2/J_1= -2 + D/(2J_1)$. Across
that line, different types of one--spin flips on the
S=1 sublattice cost lowest energy, accompanied by a change
in the sign of the slope of the magnetization of that sublattice at
low temperatures. At
fixed value of $J_2/J_1$, the range, in $D/J_1$, of compensation points 
is quite narrow, becoming extremely small for decreasing
nnn couplings, as we have shown by decreasing $J_2/J_1$ from
$-0.2$ to $-0.8$.

The transition to the paramagnetic phase appears to be always in the
universality class of the two--dimensional Ising model, with
the critical maxima of the specific heat growing with
lattice size, $L$, in a logarithmic fashion, and with
the maxima of the appropriate (staggered) sublattice susceptibilities
growing with lattice size proportionally to $L^{7/4}$. These
results hold in case of AF, 0I, and B--AF ground states
of the model. In case of the 0II ground state, there is no
phase transition.

The model is observed to exhibit interesting anomalies by
showing non--monotonic temperature dependences for
various quantities, including sublattice magnetization
and specific heat. The anomalies are typically caused by
energetically favoured flips of S= 1 spins between the state 0
and a non--zero state.

In conclusion, our study provides clear evidence for 
non--expected compensation points in a rather simple mixed Ising
model with antiferromagnetic next--nearest neighbour
couplings between S=1 spins on
a square lattice. Extensions to three dimensional lattices are
well beyond the present study, and may be investigated in the future.

\vspace{1cm}

\section {Acknowledgements}

C E thanks the Department of Physics at the RWTH Aachen
for the kind hospitality during his stay there. We thank 
Prof. Mark Novotny for a very useful discusssion as well as Dr. Mukul
Laad for interesting conversations.

\end{document}